\documentclass[a4paper,12pt]{article}
\usepackage{amsmath}
\usepackage{amssymb}
\usepackage[latin1]{inputenc}
\usepackage{amsfonts}
\usepackage{graphics}
\usepackage{epsfig}
\usepackage{epic}
\usepackage{inputenc}

\begin{document}

\renewcommand{\theequation}{\thesection.\arabic{equation}}
\thispagestyle{empty}
\vspace*{-1.5cm}
\hfill{\small KL--TH04/02}\\[8mm]

\begin{center}
{\large LIFTING A CONFORMAL FIELD THEORY\\
FROM D-DIMENSIONAL FLAT SPACE TO (D+1)-\\[1.5mm]
DIMENSIONAL ADS SPACE}\\
\vspace{2cm}
{\large W. Rühl}\\
Department of Physics, Kaiserslautern University of Technology,\\P.O.Box 3049,
67653 Kaiserslautern, Germany\\
\vspace{5cm}
\begin{abstract}
A quantum field theory on Anti-de-Sitter space can be constructed from a conformal field
theory on its boundary Minkowski space by an inversion of the holographic mapping. The 
resulting theory is defined by its Green functions and is conformally covariant. 
The structure of operator product expansions is carried over to AdS space. We show
that this method yields a higher spin field theory HS(4) from the minimal conformal
O(N) sigma model in three dimensions.
\end{abstract}
\vspace{5cm}
{\it June 2004}
\end{center}
\newpage

\section{Introduction}
AdS/CFT correspondence has so far remained a hypothesis and is connected with two
classes of models. In the first class \cite{1,2,3} an AdS(d+1) field theory is derived from
superstring theory. It possesses a large supersymmetry and the corresponding holographic image
on $\mathbf{R}(d)$ is a superconformal field theory. In Maldacena's standard case starting from type 
IIB superstrings a compactification to AdS(5)$\times$ S(5) is performed and the fields are expanded
on S(5) in harmonics (Kaluza-Klein expansion). Supergravity on AdS(5) with S(5) towers of harmonics
is mapped hypothetically on super-Yang-Mills theory SYM(4) with supersymmetry $\mathcal{N}\ = 4$
and gauge group SU(N). Perturbative expansions of AdS(5) supergravity in the Newton coupling 
constant $G_N$ go into $\frac{1}{N^{2}}$ expansions of SYM(4) at large 'tHooft coupling $\lambda$,
and only gauge invariant fields of SYM(4) appear. The harmonics on S(5) map onto representations
of the R-symmetry group SU(4) (SO(6)). The checks performed of the AdS/CFT correspondence are 
almost all based on supersymmetry. 

In the second class we have  models such as higher spin models HS(d+1) on AdS(d+1) with conformal field
theories on $\mathbf{R}(d)$ as proposed holographic images \cite{4,5}. Supersymmetry may be present or not.
The holographic map is assumed to map small-coupling expansion on small-coupling expansion. A proof of the 
correspondence can be performed only by an analysis of graphs. 

In this class an HS(4) theory with the minimal conformal O(N) sigma model as (hypothetical) holographic image
on $\mathbf{R}(3)$ is a standard case. The sigma model possesses only the single parameter $N$ and is perturbative
in $\frac 1N$ and  renormalizable. It simplifies considerably at $d = 3$ \cite{6}. The n-point function of the 
scalar ``auxiliary" field $\alpha(x)$ can be given explicitly at order $O(\frac 1N)$ for all $n$. The corresponding
model  HS(4) is based on a scalar field $\sigma(z)$  and symmetric tensor fields $h^{(l)}(z)$ for all $l \in 2\mathbf{N}$.
In the sigma model the corresponding currents $J^{(l)}$ are conserved for $l = 2$  and almost conserved (with
divergence of order $\frac 1N$ for $l \geq 4$. The anomalous conformal dimensions of $\alpha$ and $J^{(l)}$ (l $\geq $4)
are renormalization effects and known \cite{7}, and can at present not be reproduced as ``anomalous masses" of $\sigma$
and $h^{(l)}$ in HS(4). On the other hand the n-point functions of $\sigma$ can be reproduced in HS(4) (see Section 6)
by a perturbative approach.

Whereas in the sigma model the internal lines of $\alpha$ and the O(N)-vector Lorentz-scalar field $\vec \varphi$ appear,
a (still tentative) perturbative expansion of HS(4) \cite{8,9} allows internal lines only of $\sigma$ and all $h^{(l)}$.
Since the currents $J^{(l)}$ are composites of $\vec\varphi$ (bilinear at leading order), the question arises whether
to any perturbative order the exchange of such composites can simulate the exchange of $\vec\varphi$
itself. Exchanging all $J^{(l)}$  and appropriately summing over all $l$  with coupling constants adjusted,
can be shown to lead to an exchange of a bilocal biscalar O(N)-scalar field
\begin{equation}
\frac{1}{\sqrt{N}} \vec\varphi(x_1)\vec\varphi(x_2)
\label{1.1}
\end{equation}
Since at $O(\frac 1N)$ the n-point functions can be described at $d=3$ by the exchange of this bilocal field,
it is conceivable that n-point functions on AdS space of the field $\sigma$ can be described by the exchange of all
$h^{(l)}$ in a simple fashion with coupling constants adjusted as in the sigma model. 

In this work we describe a formalism to generate field theories on AdS(d+1) spaces starting from conformal field theories
on flat spaces $\mathbf{R}(d)$. The AdS theories are defined by Green functions whose boundary limits are identical 
with the CFT Green functions. Our formalism consists of two essential steps:
(1) an integral transform (''lifting") applied to all Green functions of the CFT;
(2) a representation theoretical reduction part.
The reduction applies to all external lines (technically simple) and internal lines. In the latter case we can perform
a conformal partial wave expansion and reject all unwanted representations. Instead we can also use a ``sewing procedure" of graphs
producing one internal line from two external lines belonging to a dual pair of representations by integration over the AdS space boundary. 
We hope that this construction enables us one day to formulate perturbative AdS field theories 
which are renormalizable and complete to all orders in an autonomous fashion.
 
\setcounter{equation}{0}
\section{The lifting program}
In his seminal paper ``Intertwining operator realization of the AdS/CFT 
correspondence'' V.K. Dobrev \cite {10} described the representation theoretic relation 
between a CFT on flat d-dimensional space and its partner field theory on 
AdS$_{d+1}$. Here we want to focus on the aspect of lifting a CQFT$_d$ to a 
(conformal) quantum field theory on AdS$_{d+1}$. This lifting is based on a 
technical procedure which we want to describe first.
Consider an $n$-point function
\begin{eqnarray}
G^{(n)} (x_1,...,x_n) &=& \langle \varphi_1(x_1) \varphi_2(x_2) ... \varphi_n(x_n) \rangle_{\rm CFT}  \\ \nonumber
& & x_i \in \mathbb{R}_d
\label{2.1}
\end{eqnarray}
Let the field $\varphi_k(x_k)$ be scalar, and convolute it with the dual boundary-to-bulk propagator
\begin{equation}
K_{d-\Delta_k} (z_k;x_k) = \left( \frac{z_{k0}}{z^2_{k0} + (\vec z_k -\vec x_k)^2} \right)^{d-\Delta_k}
\label{2.2}
\end{equation}
\[ \Delta_k : \mbox{conf. dimension of } \varphi_k \]
by integration over $\mathbf{R}_d$. It results an expression possessing a boundary 
asymptotic behaviour at $z_{k0} = 0$ that is not unique but splits into a sum of 
several, in simple cases (e.g. propagators) just two analytic functions behaving 
as (for propagators)
\begin{eqnarray}
z_{k0}^{\Delta_k}~ {\rm resp}~ z_{k,0}^{d-\Delta_k}  \nonumber \\
\Delta_k = \frac d2 + \left( \frac{d^2}{4} + m^2_k \right)^{\frac12}
\label{2.3}
\end{eqnarray}
We drop the second analytic function in this case, and eventually all other
functions belonging to unwanted representations.

In explicit cases this split and rejection was done by using hypergeometric 
function identities. In abstract each analytic function belongs to a different  
representation so that the splitting can be defined by a Hilbert transform 
over the Plancherel measure of the conformal group. The two representations
in (2.3) are dual to each other. Two elementary representations are dual to 
each other if their tensor type is equal and their conformal dimensions add
up to the space dimension $d$.

We proceed in this way on all $n$ legs of $G^{(n)}$ and call the result
\begin{equation}
\Gamma^{(n)}(z_1,z_2,...z_n)
\label{2.4}
\end{equation}
the ``AdS-lifted'' $n$-point function.

However, in AdS field theories one likes to introduce AdS-local interactions
implying Witten graphs whose vertices are integrated over all AdS space. Before we
can combine such construction of AdS Green functions with our lifting and reduction
technique, we must study simple Green functions in detail.

It is known that customary CFTs are built on a finite number of fundamental 
fields so that all other conformal fields are composite fields of these in a 
perturbative sense. Such theory contains 1PR graphs with fundamental fields 
at internal lines. On the other hand we can project a Green function on a 
conformal partial wave corresponding to a composite field as a unique  
representation. The 1PR graph for the exchange of the fundamental field 
contains instead a pair of dual representations: that of the fundamental 
field and its dual (or ``shadow representation'') as described above. A composite field 
is exchanged in a 1PR graph if it appears in a partial wave decomposition and
necessarily is not accompanied by the dual representation. The 
holographic image of an AdS field theory never contains the shadow 
representations. Thus we would conclude: A consistent lifting of a CFT$_d$ to 
an AdS$_{d+1}$field theory is possible only if the CFT$_d$ possesses 
solely composite fields as conformal fields. This implies that the CFT$_d$ 
is a reduced field theory of a bigger one including fundamental fields. We 
shall see, however, that the minimal O(N) sigma model contains the 
auxiliary field $\alpha$  as an elementary field and can nevertheless be lifted. 
The dual field of $\alpha$ is eliminated by a typical dynamical property of
the sigma model.
 
Let us make a further remark. Consider the case of tensor fields 
$\varphi_i(x_i)$, i.e. traceless SO(d) tensors of Young symmetry $Y_i$. 
They can be contracted (operation ``C'') to a scalar multilocal field 
in some way
\begin{equation}
C \prod^n_{i=1} \varphi_i(x_i)
\label{2.5}
\end{equation}
Then the integral
\begin{equation}
\int d \vec x C \prod^n_{i=1} \varphi (x_i-x)
\label{2.6}
\end{equation}
is a conformally invariant multilocal field if and only if the dimensions satisfy
\begin{equation}
\sum^n_{i=1} \Delta_i = d
\label{2.7}
\end{equation}
This is the background for contracting $\varphi_i(x_i)$ in $G^{(n)}$ of dimension 
$\Delta_i$ with $K_{d-\Delta_i}(z_i,x_i)$ belonging to the dual representation. 
On the other hand $K_{d-\Delta}$ intertwines a tensor $Y_i$ of SO(d) with a 
tensor $\tilde Y_i$ of SO(d+1) and $\tilde Y_i$ is not unique. The requirement 
is only that in restricting SO(d+1) to SO(d) the tensor $Y_i$ must occur. But 
there is a ``canonical'' procedure to fix this ambiguity, namely we let the Young 
tableaus of $Y_i$ and $\tilde Y_i$ be the same and both tensors traceless.

\setcounter{equation}{0}
\section{The lifting of propagators}
 Using the lifting technique we produce AdS propagators for arbitrary symmetric 
 traceless fields \cite{11,12}. Let
\begin{eqnarray}
u &=&  \zeta -1 \\ \label{3.1}
u &=& \frac{(z_1-z_2)^2}{2 z_{10}z_{20}}, \quad \zeta = \frac{z^2_{10}+z^2_{20}+ (\vec z_1- \vec z_2)^2}{2 z_{10}z_{20}}
\label{3.2}
\end{eqnarray}
be the ``chordal'' distance squared between the point $z_1,z_2$ in Poincare coordinates. 
The (canonical) bulk-to-boundary propagators for a symmetric traceless tensor 
field of rank $l$ \cite{10} is 
\begin{eqnarray}
K_\Delta^{(l)} (a,\vec b; z, \vec x) &=& \frac{z_0^{\Delta-l}}{(z^2 _0+(\vec z - 
\vec x)^l)^\Delta}  \nonumber \\
& \times & \left\{ \left(  \sum^{d+1}_{\alpha=1} \sum^d_{i=1}a_\alpha 
r(z-\vec x)_{\alpha i} b_i \right)^l - {\rm traces}  \right\} \label{3.3}
\end{eqnarray}
\begin{equation}
r(z)_{\alpha i} = 2 \frac{z_\alpha z_i}{z^2} - \delta_{\alpha i}, \quad z^2 
= z^2_0 + \vec z ^2
\label{3.4}
\end{equation}
\begin{equation}
\sum_\alpha r(z)_{\alpha j} r(z)_{\alpha i} = \delta_{ij}
\label{3.5}
\end{equation}
$a$ and $\vec{b}$ are arbitrary fixed vectors used for symmetrization. Using (3.5)
the subtraction of traces in (3.3) can be achieved by using Gegenbauer polynomials
$C_n^\lambda$, $\lambda = \frac{1}{2}(d+1) - 1$.
Contracting and integrating
\begin{equation}
\int d \vec x K_\Delta^{(l)} (a_1, \vec \nabla_b; z_1, \vec x) K^{(l)}_{d-\Delta}
 (c, \vec b; z_2,\vec x)
\label{3.6}
\end{equation}
and performing the Hilbert-transform split we can dispose of the dual term and 
get a propagator which involves Legendre functions of the second kind
\begin{eqnarray}
\Lambda_s(\zeta) = \frac{( \mu)_s}{(d-\Delta)_s} \quad (2\zeta)^{-(\Delta+s)} 
~_2F_1 ( \frac12 (\Delta + s), \frac12 ( \Delta + s +1 );  
\Delta-\mu+1; \zeta^{-2} ) \\ \label{3.7}
(\mu = \frac12 d, \zeta = \zeta(z_1, z_2)) \nonumber
\end{eqnarray}
Moreover this propagator is a bilocal tensor at $z_1, z_2$ and can be spanned by 
the algebraic basis
\begin{equation}
L_1 = \frac{(\vec a, \vec c)}{z_{10} z_{20}}, \quad L_2 = \frac{a_0c_0}{z_{10}
z_{20}}
\label{3.8}
\end{equation}
\begin{equation}
L_3 = \frac{a^2c^2_0 + a^2_0c^2}{z^2_{10}z^2_{20}}, \quad L_4 = \frac{a^2c^2}{z^2_{10}z^2_{20}}
\label{3.9}
\end{equation}

Then the final propagator is
\begin{equation}
\kappa_l(\mu,\Delta) \sum^l_{k=0} Q^{(l)}_k (\zeta) \sum_{r_1r_2r_3r_4} 
R^{(l,k)}_{r_1r_2r_3r_4} (\mu) L_1^{r_1} L_2^{r_2} L_3
\end{equation}
\label{3.10}
where $\kappa_l$ is a normalization
\begin{equation}
Q^{(l)}_k(\zeta) = \sum^{l-k}_{s=0} \frac{(-l+k)_s(k+1)_s}{(\mu+k)_s s!} (2\zeta)^s \Lambda_{s+k}(\zeta)
\label{3.11}
\end{equation}
and $R^{(l,k)}(\mu)$ are certain rational functions of $\mu$. The sum over 
$\{r_i\}$ is restricted to
\begin{equation}
r_1+r_2+2r_3+2r_4 = l
\label{3.12}
\end{equation}
The functions $R^{(l,k)}(\mu)$ are computable by a computer algorithm but we have 
no closed form for them in general.

\setcounter{equation}{0}
\section{The lifting of three-point functions}
We want to consider for simplicity scalar fields only, their dimensions are 
$\Delta_i$. In CFT$_d$ such three-point function is
\begin{equation}
f_{123} (x^2_{12})^{-h_{12}} (x^2_{23})^{-h_{23}} (x^2_{31})^{-h_{31}}
\label{4.1}
\end{equation}
with the ``coupling constant'' $f_{123}$ and
\begin{equation}
h_{ij} = \frac12 (\Delta_i + \Delta_j - \Delta_k) ~ ({\rm cycl.})
\label{4.2}
\end{equation}
In AdS$_{d+1}$ we  obtain for the ``star graph''
\begin{equation}
\Gamma^{(3)}(x_1,x_2,x_3) = \pi^{-\mu} \int \frac{du_0 d\vec u}{u_0^{d+1}} 
\prod^3_{i=1}
\left( \frac{u_0}{u^2_0 + (\vec u-\vec x_i)^2} \right) ^{\Delta_i}
\label{4.3}
\end{equation}
the same function (4.1) as in CFT with exception of the
normalization. This is an important result: Any CFT three-point function can be 
reproduced by a local interaction integral (3-star graph) in AdS field theory.
 
By lifting leg ``3'' we obtain in (\ref{3.3}) a bulk-to-bulk propagator
\begin{eqnarray}
\Gamma^{(3)}(x_1,x_2,x_3) &=& \pi^{-\mu} \int \frac{du_0 d \vec u}{u_0^{d+1}} 
\prod^2_{i=1} \left( \frac{u_0}{u^2_0 + (\vec u - \vec x_i)^2} \right)^{\Delta_i}  
\nonumber \\
& \times & \sum^\infty_{M=0} C_M (\Delta_3) \left( \frac{2u_0z_0}{u^2_0+z^2_0+
(\vec u - \vec z)^2} \right)^{\Delta_3+2M}
\label{4.4}
\end{eqnarray}
with
\begin{equation}
C_M(\Delta) = \frac{(\frac12 \Delta)_M (\frac12 (\Delta+1))_M}{M! (\Delta -
 \mu +1)_M}
\label{4.5}
\end{equation}
With standard methods the following result for $\Gamma^{(3)}$ can be derived
\begin{eqnarray}
\frac{2^{2\Delta_3-1}}{\Gamma(\Delta_3)} \sum^\infty_{M=0} \frac{1}{M!} 
\frac{\Gamma(\tau_{123}-\mu+M)\Gamma(\tau_{123}-\Delta_1+M) \Gamma(\tau_{123}-
\Delta_2+M)}{(\Delta_3-\mu+1)_M \Gamma(\tau_{123}+M)} \\ \nonumber
\times _2F_1 (\tau_{123} - \Delta_1+M, \Delta_2; \tau_{123}+M; \eta)
\label{4.6}
\end{eqnarray}
with
\begin{eqnarray}
\tau_{123} &=& \frac12 (\Delta_1+\Delta_2+\Delta_3) \\ \label{4.7}
\eta &=& 1 - \frac{(x_1-z)^2(x_2-z)^2}{x^2_{12}z^2_0}
\label{4.8}
\end{eqnarray}
Now we use the hypergeometric identity (``Kummer relation'' \cite{22} 9.132.1) 
to replace $_2F_1(\eta)$ by two different Gaussian series $_2F_1\left((1-\eta)^{-1}\right)$ which we call the 
``A-term'' and the ``B-term''.

After a few manipulations the A-term gives
\begin{eqnarray}
2^{2\Delta_3-1} \left( \prod^3_{i=1} \frac{\Gamma(\tau_{123}-\Delta_i)}{\Gamma(\Delta_i)} \right) \Gamma(\tau_{123}-\mu)  \nonumber \\
\times (x^2_{12})^{\tau_{123}-\Delta_1-\Delta_2} \left( \frac{z_0}{(x_1-z)^2} \right)^{\tau_{123}-\Delta_2} \left( \frac{z_0}{(x_2-z)^2} \right)^{\tau_{123}-\Delta_1}  \nonumber \\
\times _2F_1 (\tau_{123} - \Delta_1, \tau_{123}-\Delta_2; \Delta_3-\mu+1, (1-\eta)^{-1})
\label{4.9}
\end{eqnarray}
In the boundary limit we get
\[ \lim_{z_0\to0} z_0^{-\Delta_3} (\mbox{A-term})|_{\vec z = \vec x_3} = 
(\ref{3.1})
\]
with a certain $f_{123}$.

The B-term is instead
\begin{eqnarray}
- 2^{2\Delta_3-1} \frac{\Gamma(\Delta_3-\mu+1)}{(\tau_{123}-\mu)(\tau_{123}-
\Delta_3) \Gamma(\Delta_3)} 
 \left( \frac{z_0}{(x_1-z)^2} \right)^{\Delta_1} \left( 
\frac{z_0}{(x_2-z)^2} \right)^{\Delta_2} \nonumber \\ 
\times _3F_2(1, \Delta_1, \Delta_2; 
\tau_{123} - \Delta_3+1, \tau_{123}-\mu+1; 
(1-\eta)^{-1} )
\label{4.10}
\end{eqnarray}
If
\begin{equation}
\Delta_1 + \Delta_2 < \Delta_3
\label{4.11}
\end{equation}
the B-term dominates the A-term on the AdS boundary, and it results
\begin{eqnarray}
\lim_{z_0\to 0} z_0^{-(\Delta_1+\Delta_2)} (\mbox{B-term})|_{\vec z = \vec x_3} 
\nonumber \\
= {\rm const}~ (\vec x^2_{13})^{-\Delta_1} (\vec x^2_{23})^{-\Delta_2}
\label{4.12}
\end{eqnarray}
i.e. the 3-point function of $\varphi_1(\vec x_1), \varphi_2(\vec x_2)$ and 
$\varphi_1\varphi_2(\vec x_3)$ in CFT$_d$. 

Thus the B-term couples the representations $[\Delta_1, 0]$ and $[\Delta_2, 0]$
to an infinite number of representations which are symmetric traceless tensors
of rank $l$ and conformal dimension
\begin{equation}
\Delta = \Delta_1 + \Delta_2 + l + 2t  (t \in \mathbf{N_0})
\label{4.13}
\end{equation}
These are unwanted and we eliminate them by just rejecting the whole B-term.

It is of course desirable to extend this result to 3-point functions of 
(symmetric traceless rank $l_i$) tensor fields with dimensions
\begin{equation}
\Delta_i = \Delta_{i0} + l_i
\label{4.14}
\end{equation}
In this case the ranks $l_i$ must satisfy obviously (for arbitrary $i,j,k$ in cyclic order)
\begin{equation}
|l_i-l_j| \le l_k \le l_i+l_j,
\label{4.15}
\end{equation}
to enable a contraction of the tensors. If (4.15) is satisfied, a unique AdS local 
interaction is possible. Then the B-term
contains composite fields in the product representation of $[\Delta_1,l_1]$
and $[\Delta_2,l_2]$. Also these B-terms are discarded.

\setcounter{equation}{0}
\section{Operator product expansions for AdS$_{d+1}$ field theories lifted from a 
CFT$_d$}
An AdS$_{d+1}$ field theory obtained by lifting from a conformal field theory in 
flat d-dimensional space satisfying all the constraints found, inherits the 
algebraic structure of OPEs. Since the bulk-to-boundary propagators are 
intertwiners for conformal groups, it is a conformally covariant field theory. We 
shall formulate it here in a non-group theoretical fashion.

We can express the OPE structure of a conformal field theory by 3- and 4-point 
functions. Namely, given a CFT$_d$ based on certain finitely many fundamental 
fields which transform as ``elementary'' representations of the conformal group 
(in short: are ``conformal''), then each pair of fundamental fields expands as
\begin{equation}
\varphi_i(x_i) \varphi_j(x_j) = \sum_k \int K_{ijk}(x_i,x_j,x) \phi_k(x) d^dx
\label{5.1}
\end{equation}
where $\phi_k$ are conformal fields (fundamental or composite) and $K$ are 1-leg 
amputated 3-point  functions. This expansion is ``weakly convergent'' and can be 
extended to all conformal fields (not only fundamental fields to start with).

Instead we can consider 4-point functions
\begin{eqnarray}
\langle \varphi_1(x_1) \varphi_2(x_2) \varphi_3(x_3) \varphi_4(x_4) \rangle 
\nonumber \\
= \sum_k \int d^dx d^dy ~ K_{12k}(x_1,x_2;x) G^{(2)}_k(x,y) K_{34k}(x_3,x_4;y)
\label{5.2}
\end{eqnarray}
where $G_k^{(2)}(x,y)$ is the two-point function of the field $\phi_k$.
Thus we decompose the 4-point function (in a convergent fashion)
into a sum of exchange amplitudes (= `` conformal partial waves''), corresponding 
to graphs

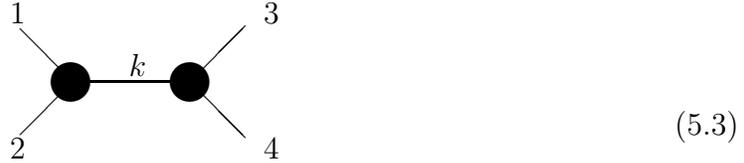
\begin{align}
\begin{picture}(40,50)
\put(-24,0){\line(1,1){20}} \put(-24,40){\line(1,-1){20}}
\put(40,20){\line(1,1){21}}
\put(40,20){\line(1,-1){21}}\put(40,20){\line(-1,0){44}}
\put(17,22){$k$} \put(-28,42){$1$} \put(-28,-9){$2$}
\put(68,42){$3$} \put(68,-9){$4$}
\put(-5,20){\circle*{20}}\put(40,20){\circle*{20}}
\end{picture}
\end{align}\vspace{1cm}

the 3-point function subgraphs being nonlocal in general. The
identification of each exchange amplitude with a ``conformal
partial wave'' i.e. an irreducible elementary representation of
the conformal group is nontrivial. The exchange amplitudes are
easy to calculate and known explicitly for all exchanges of tensor
fields between four scalar fields \cite{7}.

We would like to use exchange amplitudes for the operator product expansions in 
the AdS$_{d+1}$ field theories as well. Thus we study these in a systematic way, 
allowing us to impose the requirements of ``irreducibility'' of the exchanged 
object. Here we want to concentrate on the case of a scalar field exchange. Tensor 
fields can be treated analogously. The well-known work by Liu \cite{13} cannot be 
used since the integrations involved contain a systematic error.

We start from the singly lifted 3-point function of Section 3 and
join the two boundary-to-bulk propagators for the fields $\phi_4,
\phi_5$ 
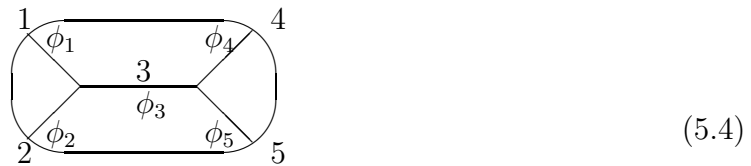
\vspace{1cm}\begin{align}
\begin{picture}(40,50)
\put(20,20){{\oval(100,50)}} \put(-24,0){\line(1,1){20}}
\put(-24,40){\line(1,-1){20}} \put(40,20){\line(1,1){21}}
\put(40,20){\line(1,-1){21}}\put(40,20){\line(-1,0){44}}
\put(17,22){$3$} \put(-28,42){$1$} \put(-28,-9){$2$}
\put(68,42){$4$} \put(68,-9){$5$}
\put(43,-1){$\phi_{5}$}\put(43,35){$\phi_{4}$}
\put(-17,-1){$\phi_{2}$}\put(-17,35){$\phi_{1}$}\put(17,10){$\phi_{3}$}
\end{picture}
\end{align}\vspace{1cm}

We decompose the $\phi_1\phi_2\phi_3$ 3-point function into A- and
B-term. Then we expand the $_2F_1$ resp $_3F_2$ functions into
power series. Each power leads by integration over the vertex
``3'' to an AdS four-star function which is well-known
\cite{13,14,15}. It can be decomposed into two Gausssian
hypergeometric series. So we obtain four expressions
\[ {\rm A} \to {\rm A}_1 +{\rm A}_2 \]
\[ {\rm B} \to {\rm B}_1 + {\rm B}_2 \]
But we can show that A$_2$ + B$_2$ can be summed to one expression C$_1$. These 
functions have the properties: \\[2mm]
A$_1$: describes the exchange of the field $\phi_3$ in an irreducible fashion; 
\\[2mm]
B$_1$: describes the exchange of all conformal composite fields of $\phi_1$ and 
$\phi_2$ of dimensions
\begin{equation}
\Delta_1 + \Delta_2 + l_{12} + 2t_{12}; (l_{12}, t_{12} \in \mathbf{N}_0)
\label{5.3}
\end{equation}
C$_1$: describes the exchange of all conformal composite fields of $\phi_4$ and 
$\phi_5$ of dimensions
\begin{equation}
\Delta_4 + \Delta_5 + l_{45} + 2t_{45}; (l_{45}, t_{45}  \in \mathbf{N}_0)
\label{5.4}
\end{equation}
and is obtained from B$_1$ by reflection
\[ 1 \leftrightarrow 4 \]
\[ 2 \leftrightarrow 5 \]
Here $l_{ij}$ is the rank of a symmetric traceless tensor and $l_{ij} + 2t_{ij}$ 
is the number of derivations applied to bulk fields, of which $t_{ij}$ pairs are
contracted. $l$ and $t$ fix the conformal composite field uniquely.

The A$_1$ amplitude is the AdS-conformal partial wave and differs from the 
corresponding CFT partial wave by the normalization. The B$_1$ and C$_1$ 
amplitudes are typical AdS field theory artefacts.

The explicit forms of these A$_1$,B$_1$,C$_1$ amplitudes are given by 
hypergeometric 2-variable series with the variables $u$ and $1-v$
\begin{equation}
u = \frac{x^2_{12}x^2_{45}}{x^2_{14}x^2_{25}}, \quad v = \frac{x^2_{15}x ^2_{24}}
{x^2_{14}x^2_{25}}
\label{5.5}
\end{equation}
We integrate the $x_3$ vertex by
\begin{equation}
\pi^{-\mu} \int \frac{dx_{30}d \vec x_3}{x_{30}^{d+1}}
\label{5.6}
\end{equation}
and obtain after some summations
\begin{eqnarray}
\Gamma^A_{\phi_1\phi_2 \to \phi_3 \to \phi_4\phi_5} &=& N(A)(x^2_{14})^{-\frac12 (\Delta_3+\Delta_4-\Delta_5)} (x^2_{15})^
{-\frac12(\Delta_1+\Delta_5-\Delta_2-\Delta_4)} \nonumber \\
& \times & (x^2_{25})^{-\frac12 (\Delta_3-\Delta_1+\Delta_2)} (x^2_{45})^
{-\frac12(-\Delta_3+\Delta_4+\Delta_5)} (x^2_{12})^{-\frac12 (\Delta_1+\Delta_2-\Delta_3)} \nonumber \\
&\times & \sum^\infty_{n,m=0} a_{nm} \frac{u^n(1-v)^m}{n!m!}
\label{5.7}
\end{eqnarray}
where
\begin{eqnarray}
a_{nm} &=& \frac{(\frac12(\Delta_3+\Delta_1-\Delta_2))_n (\frac12( \Delta_3-
Delta_4+\Delta_5))_n (\frac12(\Delta_3-\Delta_1+\Delta_2))_{n+m}}
{(\Delta_3)_{2n+m} (\Delta_3-\mu+1)_n} \nonumber \\
& \times & (\frac12 (\Delta_3+\Delta_4-\Delta_5))_{n+m}
\label{5.8}
\end{eqnarray}
and
\begin{eqnarray}
N(A) &=& \frac{2^{2\Delta_3-1} \Gamma(\tau_{123}-\mu) \Gamma(\tau_{345}-\mu)}
{(\prod^5_{i=1} \Gamma(\Delta_i)) \Gamma(\Delta_3) \Gamma (\Delta_3-\mu+1)}
\nonumber \\
& \times & \prod^3_{i=1} \Gamma(\tau_{123}-\Delta_i) \prod^5_{j=3} \Gamma(\tau_
{345}-\Delta_j)
\label{5.9}
\end{eqnarray}
We use the shorthand
\begin{eqnarray}
\tau_{123} &=& \frac12 (\Delta_1+\Delta_2+\Delta_3) \label{5.10} \\
\tau_{345} &=& \frac12 (\Delta_3 + \Delta_4 + \Delta_5)
\label{5.11}
\end{eqnarray}
In a similar way we obtain the B$_1$ amplitude
\begin{eqnarray}
\Gamma^B_{\phi_1\phi_2 \to \phi_3 \to \phi_4\phi_5} &=& N(B)(x^2_{14})^{-\frac12 (\Delta_3+\Delta_4-\Delta_5)} (x^2_{15})^{-
\frac12(\Delta_1+\Delta_5-\Delta_2-\Delta_4)} \nonumber \\
& \times & (x^2_{25})^{-\frac12 (\Delta_3-\Delta_1+\Delta_2)} (x^2_{45})^{-
\frac12(-\Delta_3+\Delta_4+\Delta_5)}(x^2_{12})^{-\frac12 (\Delta_1+\Delta_2-\Delta_3)} u^{\tau_{123}-
\Delta_3} \nonumber \\
& \times & \sum^\infty_{n,m=0} \frac{u^n(1-v)^m}{n!m!} b_{nm}
\label{5.12}
\end{eqnarray}
where
\begin{eqnarray}
b_{nm} &=& \frac{(\Delta_1)_n (\tau-\Delta_4)_n (\Delta_2)_{n+m}(\tau-\Delta_5)_
{n+m}}
{(\Delta_1+\Delta_2)_{2n+m} (\tau_{123}-\tau_{345}+1)_n} \nonumber \\
& \times & _3F_2(\tau-\mu,-n,1;~ \tau_{123}-\Delta_3+1, \tau_{123} - \mu+1;1)
\label{5.13}
\end{eqnarray}
and
\begin{equation}
\tau = \frac12 (\Delta_1+\Delta_2+\Delta_4+\Delta_5)
\label{5.14}
\end{equation}
and
\begin{eqnarray}
N(B)&=& - \frac{2^{2\Delta_3-2}}{\Gamma(\Delta_1+\Delta_2) \Gamma(\Delta_3)^2} 
\prod_{i \in \{4,5\}} \frac{\Gamma(\tau-\Delta_i)}{\Gamma(\Delta_i)} \nonumber \\
& \times & \frac{\Gamma(\tau-\mu) \Gamma(\Delta_3-\mu+1)}
{(\tau_{123}-\mu)(\tau_{123}-\Delta_3)}
\label{5.15}
\end{eqnarray}

In the case of an exchange of a tensorial conformal field of dimension $\Delta_3$, 
the A-term can be obtained as follows. We evaluate the AdS-3-point functions of 
two scalar fields 1,2 with the tensor field of dimension $\Delta_3$ and the two 
scalar fields 4,5 with the same tensor field but dimension $d-\Delta_3$ (for this 
integration, see \cite{7,16}). Both tensorial legs are drawn to the boundary point 
$x_3$, contracted, and integrated over $x_3$. If necessary the dual representation
term is discarded. On the one hand we obtain by the arguments of Section 3 
that the two vertices have been connected this way by a bulk-to-bulk propagator 
of the tensor field. On the other hand AdS 3-point functions with all legs on the 
boundary are equal to flat CFT 3-point functions, the integration leads to the CFT 
exchange amplitude \cite{7,16}, and the exchanged object is irreducible, namely 
only the tensor field representation. Thus the B$_1$ and C$_1$-part of the 
exchange amplitude have dropped out automatically.

This sewing of two AdS three-point functions to an exchange graph four-point
function can be generalized to any pair of AdS graphs. Essential is that in
one graph an external line belonging to a representation $[\Delta, l]$
corresponds to an external line of the other graph with dual representation 
$[d-\Delta,l]$, that both external lines are extended to a common point $x$ on 
the AdS boundary, and that this point is integrated over. It results one graph 
with an internal line carrying a dual pair of representations. One of these has to
be discarded if the line corresponds to a propagator of a composite field.

We conclude that for AdS 4-point functions with all external lines on the boundary
operator product expansions are identical with those of flat CFT. Only
proportionality factors (``coupling constants'') must be adjusted.

\setcounter{equation}{0}
\section{Lifting the minimal conformal O(N) sigma model in d=3 to the higher spin 
field theory HS(4) on AdS$_4$}
In a conformal O(N) sigma model the O(N)-vector Lorentz-scalar field $\vec 
\varphi(x)$ can be contracted to a bilocal O(N) singlet field
\begin{equation}
b(x,y) = \frac{1}{\sqrt N} \vec \varphi(x) \vec \varphi(y)
\label{6.1}
\end{equation}
so that
\begin{eqnarray}
\langle b(x_1,x_2) b(x_3,x_4)\rangle_{\rm CFT} = (x^2_{13} x^2_{24})^{-\delta} + (x^2_{14}x^2_{23})^{-\delta} \nonumber \\
+ O\left( \frac 1N \right) ~ {\rm corrections}
\label{6.2}
\end{eqnarray}
$\vec \varphi(x)$ has the conformal dimension
\begin{equation}
\delta = \mu - 1 + O(\frac1N), \mu = \frac12 d
\label{6.3}
\end{equation}
We shall first consider the case of a free field $\vec\varphi(x)$.

The operator product expansion of $b(x_1,x_2)$ in the free case involves the currents
\begin{eqnarray}
J^{(l)}(y;a) &=& \frac{1}{\sqrt N} \sum^{[l/2]}_{r=0} \sum^{l-2r}_{n=0}A^{(l)}_{rn}
(a^2)^r (\vec a \cdot \vec \partial)^{n} (\vec \partial)^r_\otimes \varphi_i(y) \nonumber \\
& & (\vec a \cdot \vec \partial)^{l-n-2r} (\vec \partial)^r_\otimes \varphi^i(y)
\label{6.4} \\
A^{(l)}_{rn} &=& \frac{(-1)^nl!}{2^rr!n!(l-n-2r)!} ~\frac{(\delta)_{l-r}}{(\delta)_{r+n}(\delta)_{l-n-r}} 
\label{6.5} 
\end{eqnarray}
which are conserved traceless and have dimension
\begin{equation}
\Delta^{(l)} = d - 2 + l
\label{6.6}
\end{equation}

If we decompose $J^{(l)}$ into
\begin{equation}
J^{(l)} = J^{(l)}_0 + J^{(l)}_{\rm trace}
\label{6.7}
\end{equation}
where $J^{(l)}_0$ contains all terms in (\ref{6.4}) with $r=0$, then we can show that
\begin{eqnarray}
b(x_1,x_2) &=& \sum^\infty_{M=0} \sum^M_{l=0} B_{M,l} (a \partial_y)^{M-l} J_0^{(l)} (y;a) \nonumber \\
& & (M,l~ {\rm even})
\label{6.8}
\end{eqnarray}
is a reordered Taylor expansion with
\begin{equation}
a = x_1-y = y -x_2
\label{6.9}
\end{equation}
The coefficients $B_{M,l}$ can be calculated from the linear system of equations
\begin{equation}
{\cal A}_{Mk} = \sum^M_{l=k} B_{Ml} {\cal M}_{lk}
\label{6.10}
\end{equation}
with
\begin{equation}
{\cal M}_{lk} = \left( {l \atop k} \right)\frac{(2\delta+l-1)_k}{(\delta)_k}  
\quad
\footnotemark[1]
\label{6.11}
\end{equation}
\footnotetext[1]{We define the Pochhammer symbol (O)$_0$ to be one. Then ${\cal M}_
{00} = 1$ for $d = 3$. Instead (\ref{6.13}) gives ${\cal M}_{00}= \frac12$. 
But $l=0$ is not relevant.}

\begin{equation}
{\cal A}_{Mk} = \frac{2^k}{M!} \left( {M \atop k} \right)
\label{6.12}
\end{equation}
Since ${\cal M}$ is triangular and
\begin{equation}
{\cal M}_{ll}  = 2^{2l-1}
\label{6.13}
\end{equation}
the $B_{Ml}$ are uniquely defined.

However, (\ref{6.8}) is not yet an operator product expansion since $J^{(l)}_0$
is not an irreducible representation of the conformal group. Taking this into
account we can write (\ref{6.8}) as
\begin{equation}
b(x_1,x_2)  =  \sum^\infty_{M=0} \sum^M_{l=0} [B_{M,l}(a \partial_y)^{M-l} 
J^{(l)} (y;a)) + \sum^\infty_{t=1}B_{Mlt}(a^2)^{t}(a\partial_y)^{M-l} J^{(l,t)}(y;a)]
\label{6.14}
\end{equation}
where the ``twist currents'' $J^{(l,t)}$ are traceless symmetric tensors of rank $l$
and dimension                                                                       
\begin {equation}
\Delta^{(l,t)} = 2\mu - 2 + l + 2t
\label {6.15}
\end{equation}
The simplest one is the scalar density $J^{(0,1)}$
\begin{equation}
J^{(0,1)} =  N^{-1/2}\left(\vec \partial \vec \varphi(x)\right)^2
\label{6.16}
\end{equation}
which is the Lagrangian density and conformal. It appears in (\ref{6.14}) with 
the coefficient
\begin{equation}
B_{001} = \frac{1}{2(2\delta+1)}
\label{6.17}
\end{equation}

By conformal invariance we have
\begin{eqnarray}
& &\langle J^{(l)} (y_1;a) J^{(l)} (y_2;b) \rangle_{\rm CFT} \nonumber \\
&= & {\cal N}_l (y^2_{12})^{-(d-2+l)} \nonumber \\
& &\left\{ \left[ 2 \frac{(a\cdot y_{12})(b \cdot y_{12})}{y^2_{12}} - 
(a \cdot b)\right]^l - {\rm traces} \right\}
\label{6.18}
\end{eqnarray}
and
\begin{eqnarray}
& &\langle J^{(l,t)} (y_1;a) J^{(l,t)} (y_2;b) \rangle_{\rm CFT} \nonumber \\
&=& {\cal N}_{l,t} (y^2_{12})^{-(d-2+l+2t)} \nonumber \\
& &\left\{ \left[ 2 \frac{(a\cdot y_{12})(b \cdot y_{12})}{y^2_{12}} - 
(a \cdot b)\right]^l - {\rm traces} \right\}
\label{6.19}
\end{eqnarray}
From the explicit form (\ref{6.4}) we obtain
\begin{equation}
{\cal N}_l = 2^{l+1}(l!)^2 \frac{(2\delta + l-1)_l}{(\delta)_l}
\label{6.20}
\end{equation}
and from (\ref{6.10}) - (\ref{6.13}) for comparison
\begin{equation}
B_{ll} = \frac{2^l}{l!}~ \frac{(\delta)_l}{(2 \delta+l-1)_l}
\label{6.21}
\end{equation}
We note also that the curly brackets in (\ref{6.18}), (\ref{6.19}) can be 
represented as Gegenbauer polynomials
\begin{equation}
\{ ... \} = \frac{l!}{2 ^l(\delta)_l} \eta^l C^\delta_l(\xi)
\label{6.22}
\end{equation}
with
\begin{eqnarray}
\xi &=& \frac{1}{|a||b|} ~ \left( 2 \frac{(a\cdot y_{12})(b\cdot y_{12})}{y^2_{12}} - 
(a \cdot b)\right) \label{6.23} \\[2mm]
\eta &=& |a| |b|
\label{6.24}
\end{eqnarray}

Now we investigate the corresponding bilocal biscalar field on AdS space. For this 
purpose we study the interacting case only. From a conformal partial wave expansion of 
the four-point function
\begin{equation}
\langle b(x_1,x_3), b(x_2,x_4) \rangle_{\rm CFT} \nonumber 
\end{equation}
we know that the operator product expansion of $b(x_1,x_3)$ contains currents
$J^{(l)}, J^{(l,t)}$ and the field $\alpha(x)$ 
\begin{eqnarray}
b(x_1, x_3) &=& \sum_{l=2}^{\infty}g_l \int\mathrm{d}wK^{(l)}(x_1,x_3,w)^{\mu_1\ldots\mu_l}
J^{(l)}(w)_{\mu_1\ldots\mu_l} \nonumber\\
&+& \sum_{l=0}^{\infty}\sum_{t=1}^{\infty}g_{l,t}\int\mathrm{d}wK^{(l,t)}(x_1,x_3,w)^{\mu_1
\ldots \mu_l} 
J^{(l,t)}(w)_{\mu_1\ldots\mu_l} \nonumber\\
&+& g_{\alpha} \int\mathrm{d}w K_{\alpha} (x_1,x_3,w)\alpha(w) + O(\frac{1}{N})
\label{6.25}
\end{eqnarray}                                                                            	
Note that in contrast to (\ref{6.14}) the current $J^{(0)}$ has been replaced by the $\alpha$
field. To leading order we have
\begin{eqnarray}
g_{\alpha} &=&  z^{\frac{1}{2}}_1\\
\label{6.26}
K_{\alpha}(x_1,x_3,x_5) &=& (x_{15}^{2}x_{35}^{2})^{-\delta}
\label{6.27}
\end{eqnarray}
For $K^{(l)}$ , $K^{(l,t)}$ we can make an ansatz to leading order                          
\begin{eqnarray}
K^{(l)}(x_1,x_3,x_5)^{\mu_1 \ldots \mu_l} &=& (x_{15}^{2} x_{35}^{2})^{\frac{1}{2}
(l-2)} (x_{13}^2)^{1-\delta-\frac{l}{2}} \nonumber \\
& &\left\{\bigotimes_{i=1}^{l}\frac{\eta^{\mu_i}}{|\eta|} - traces \right\}
\label{6.28}
\end{eqnarray}
with
\begin{equation}
\eta =  \frac{x_{15}}{x_{15}^2} - \frac{x_{35}}{x_{35}^2}
\label{6.29}
\end{equation}
and for $K^{(l,t)}$ the same way. Since we know that the inverse kernels of $K^{(l)}$
and $K^{(l,t)}$ are differential operators, we expect these kernels themselves to be 
regularized integral kernels.

The coupling constants $g_l$ can be calculated using the conformal partial wave decomposition
of the CFT four-point function quoted above and the free field expressions valid at
leading order
\begin{eqnarray}
 \langle J^{(l)}(x_5)_{\mu_1 \ldots\mu_l}b(x_2,x_4)\rangle_{\rm CFT} &=& 
2^{l+1}( \delta)_l (x_{25}^2 x_{45}^2)^{- (\delta + \frac{l}{2})} (x_{24}^2)^{\frac{l}{2}}
\nonumber\\
& &\left\{\bigotimes_{i=1}^{l} \frac{\xi_{\mu_i}}{|\xi|} - traces\right\}
\label{6.30}
\end{eqnarray}
Here $\xi$ is defined by
\begin{equation}
\xi  =  \frac{x_{25}}{x_{25}^2} -  \frac{x_{45}}{x_{45}^2}
\label{6.31}
\end{equation}
Since $K^{(l)}$ is obtained from (\ref{6.30}) by amputation on $x_5$, the contraction of
these two kernels and their integration over $x_5$ has been performed in the derivation of 
the ``master equation'' in \cite{7}, eqns (2.12) - (2.20). The exchanged current $J^{(l)}$
is to leading order conserved, for $l = 2$ this is the conserved energy-momentum tensor.
This necessitates a regularization of the kernels (\ref{6.30}), (\ref{6.28}).

On the other hand the Green function of two $b$-fields can be decomposed in partial waves as (see \cite{7})
\begin{eqnarray}
 \langle b(x_1,x_3), b(x_2,x_4) \rangle &=& \sum_{l=2, even}^{\infty} \gamma_{l}^{2} (x_{12}^2 x_{34}^2)^{-\delta}
 \sum_{n,m = 0}^{\infty} \frac{u^n (1-v)^m}{n!m!} a_{nm}^{(l)} \nonumber\\
&+& \sum_{l=0}^{\infty}\sum_{t=1}^{\infty}
\gamma_{l,t}^{2} (x_{12}^2 x_{34}^2)^{-\delta} u^{t} \sum_{n,m=0}^{\infty} \frac{u^n (1-v)^m}
{n!m!} a_{nm}^{(l,t)} \nonumber\\
&+& z_1^\frac{1}{2} (x_{12}^2x_{34}^2)^{-\delta} u^{\mu-2\delta}\sum_{n,m =0}^{\infty} \frac{u^n(1-v)^m}
{n!m!} \frac {(n!)^2((n+m)!)^2}{(2n+m+1)!(3-\mu)_n} \nonumber\\
&+& O(\frac{1}{N})
\label{6.32}
\end{eqnarray}
The square of the critical coupling constant is $z$ which can be expanded as
\begin{eqnarray}
z &=& \sum_{k=1}^{\infty} \frac{z_k}{N^k} \nonumber\\
z_1|_{d=3} &=& \frac{1}{\pi^{4}}
\label{6.33}
\end{eqnarray}
and $u$ and $v$ are as in (\ref{5.5}) with labels (1,2,4,5) replaced by (1,3,2,4).
Here we normalize $a_{0,l}^{(l)}$ and $a_{0,l}^{(l,t)}$ to
\begin{equation}
a_{0,l}^{(l)} = a_{0,l}^{(l,t)} = 1
\label{6.34}
\end{equation}

For the conserved currents $J^{(l)}$ the coefficients are easy to calculate
\begin{eqnarray}
a_{nm}^{(l)} &=& \sum_{s=0}^{n} (-1)^s{ n\choose s }{m+n+s \choose l} \frac{(\delta +
l)_{m-l+n}(\delta + l)_{m-l+n+s}}{(2\delta + 2l)_{m-l+n+s}}
\label{6.35}
\end{eqnarray}
The coupling constants $\gamma_{l}^2$ are also known \cite{7,5}
\begin{eqnarray}
\gamma_{l}^{2} &=& \left\{ \begin{array}{cc}
                         \frac{ 2((\delta)_l)^2}{(2\delta + l - 1)_l } & \mbox{for}\; l > 0,
			 \mbox{even} \\
                          0 & \mbox{for all other $l$}
			   \end{array} \right.  
\label{6.36}		    
\end{eqnarray}		    
From the normalization of the``master formula'' \cite{7,16} we obtain then the relation
\begin{eqnarray}
\gamma_{l}^{2}	&=&  g_{l} 2\pi^{\mu} l! (\delta)_{l} \frac{\Gamma(2-\mu+l)\Gamma(\delta+l)
^2}{\Gamma(2\delta+2l)} \\
\label{6.37}	    	   
g_l &=& \frac{\Gamma (2\delta +l -1) (2\delta +2l -1)}{\pi^{\mu}l!\Gamma(\delta)^2 \Gamma (1-\delta)}
\label{6.38}
\end{eqnarray}
which at $d = 3$ reduces to
\begin{equation}
g_l  =  \frac {2}{\pi^3}
\label{6.39}
\end{equation}
The coupling constants $g_{l,t}$ could be calculated in the same fashion (with more effort) 
if $J^{(l,t)}$ were known as a free-field expression to leading order.

Now we extend the operator product expansion (\ref{6.25}) to AdS fields
\begin{eqnarray}
B(z_1,z_3)  & = & \sum_{l=2, even}^{\infty} g_l \int\mathrm{d}w V^{(l)}(z_1,z_3,
w)^{a_1\ldots a_l} h_{a_1\ldots a_l}^{(l)}(w) \nonumber\\
&+& \sum_{l=0, even}^{\infty}\sum_{t=1}^{\infty}g_{l,t}\int\mathrm{d}w V^{(l,t)}(z_1,z_3,w)^
{a_1\ldots a_l} h_{a_1\ldots a_l}^{(l,t)}(w) \nonumber\\
&+& z_1^{\frac{1}{2}} \int \mathrm{d}w V_{\alpha}(z_1,z_3,w) \sigma(w)+ O(\frac{1}{N})
\label{6.40}
\end{eqnarray}
where the measure is on AdS
\begin{equation}
\mathrm {d}w  =  \frac{\mathrm{d}\vec{w}\mathrm{d}w_0}{w_0^{d+1}}
\label{6.41}
\end{equation}
The label $a_i$ runs over
\begin{displaymath}
\{0,1\ldots d\}
\end{displaymath}
and $h^{(l)}, h^{(l,t)}$ are symmetric traceless parts of tensor fields.

The kernels $V^{(l)}, V^{(l,t)}, V_\alpha$ are obtained by first lifting the corresponding
kernels $K^{(l)}, K^{(l,t)}$ and $K_{\alpha}$ in their first two arguments. Moreover we make
the ansatz
\begin{eqnarray}
J^{(l)}(x)_{\mu_1\ldots\mu_l} & = & 
\int\mathrm{d}zS^{(l)}(x,z)_{\mu_1\ldots \mu_l}^{a_1 \ldots a_l} h^{(l)}(z)_{a_1\ldots a_l}
\label{6.42}
\end{eqnarray}
An analogous ansatz holds for $J^{(l,t)}$ but for the $\alpha$ field we assume
\begin{equation}
\alpha(x)  = \int\mathrm{d}zS_{\beta}(x,z)\sigma (z)
\label{6.43}
\end{equation}
Here $\beta$ is the conformal dimension of $\alpha(x)$
\begin{equation}
\beta  = 2 - 2\eta(\varphi) -  2\kappa
\label{6.44}
\end{equation}
$\eta(\varphi)$ is the anomalous part of the conformal dimension of the field $\vec\varphi$
\begin{eqnarray}  
\delta &=& \mu - 1 + \eta(\phi) \\
\label{6.45}
\eta(\varphi) &=&\sum_{k=1}^{\infty}\frac{\eta_{k}(\varphi)}{N^k} \\
\label{6.46}
\eta_1(\varphi) &=&\frac{4}{3\pi^2} \quad \mbox{at} \; d = 3 \\
\label{6.47}
\kappa &=&\sum_{k=1}^{\infty}\frac{\kappa_{k}}{N^k} \\
\label{6.48}
\kappa_{1} &=& \frac{4}{\pi^2} \quad \mbox{at} \; d= 3 
\label{6.49}
\end{eqnarray}

The kernels S can be obtained from Dobrev's boundary-to-bulk propagators \cite{10} by a regularized  			   
inversion. For example in the scalar case we make the ansatz
\begin{equation}
S_{\Delta}(\vec{x},z) = lim_{\epsilon\searrow 0} p_{\epsilon}(\Delta,0)^{-1}
              \frac{z_{0}^{d + \epsilon -\Delta}}{(z_{0}^{2} + (\vec{z} - \vec{x})^2)^
	      {d-\delta}}
\label{6.50}
\end{equation}
The integral
\begin{equation}
\int\mathrm{d}z\frac{z_{o}^{d+\epsilon
-\Delta}}{(z_{o}^{2}+(\vec{z}-\vec{x}_1)^2)^{d-\Delta}}\frac{z_{0}^\Delta}{(z_{0}^2 + (\vec{z}-
\vec{x}_2 )^2)^\Delta}
\label{6.51}
\end{equation}
gives for small $\epsilon$
\begin{equation}
\pi^\mu \frac{\Gamma(\mu)\Gamma(\Delta - \mu)\Gamma(\mu-\Delta)}{\Gamma(\Delta)\Gamma(d-
\Delta)}(\vec{x}_{12}^{2})^{-\mu + \frac{1}{2}\epsilon}(1+O(\epsilon))
\label{6.52}
\end{equation}
Moreover we use
\begin{equation}
\frac{1}{\Gamma(\epsilon)}(\vec{x}_{12}^{2})^{-\mu + \frac{1}{2} \epsilon} =
\frac{2\pi^{\mu}}{\Gamma(\mu)}\delta(\vec{x}_{12}) + O(\epsilon)
\label{6.53}
\end{equation}
so that with
\begin{equation}
p_{\epsilon}(\Delta,0) = \pi^{2\mu}\frac{\Gamma(\Delta-\mu)\Gamma(\mu-\Delta)}{\Gamma(
\Delta)\Gamma(d-\Delta)} \Gamma(\epsilon)
\label{6.54}
\end{equation}
it is guaranteed that by this regularization in the limit $\epsilon \searrow 0$
\begin{equation}
\int\mathrm{d}z S_{\Delta}(\vec{x}_1, z) K_{d-\Delta}(z,\vec{x}
_{2}) = \delta(\vec{x}_{12})
\label{6.55}
\end{equation}
The function $p_{\epsilon}(\Delta,0)$ is well known from the inversion of the scalar field
propagator in CFT. In the same way we can define a function $p_{\epsilon}(\Delta, l)$ 
from the tensor field propagator.

For symmetric traceless tensors of rank l (with $\vec{\xi}, \vec{\eta}$
arbitrary vectors of $\mathbf{R}_d$ ) and after contraction of tensor labels of
$\mathbf{R}_{d+1}$ we have
\begin{eqnarray}
\int\frac{\mathrm{d}z_0\mathrm{d}\vec{z}}{z_{0}^{d+1}} S_{[\Delta,l]}(\vec{x}_1, z; \vec{\xi})
K_{[\Delta, l]}(z,,\vec{x}_2;\vec\eta)
 &=& \delta(\vec{x}_{12})[ (\vec{\xi} \cdot \vec{\eta})^{l} - traces ]
\label{6.56}
\end{eqnarray}
with a corresponding regularization.

\setcounter{equation}{0} 
\section {Application to the n-point functions of the scalar field $\sigma$ in HS(4)}
We want to study four-point functions of $\sigma$
\begin{equation}
\langle\prod_{i=1}^{4}\sigma(z_i)\rangle _{AdS}
\label{7.1}
\end{equation}
in the interacting (critical conformal) case. At leading order we obtain three disconnected
terms
\begin{equation}
\sum_{i=2}^{4} \Lambda_0(\zeta(z_1,z_i)) \Lambda_0(\zeta(z_j, z_k))
\label{7.2}
\end{equation}
where $(i,j,k)$ are cyclic permutations of $(2,3,4)$.

At the next order we expect from HS(4) all possible exchanges of the "gauge fields"
$h^{(l)}$ between pairs of $\sigma$ fields in all three channels. There is no local
triple interaction of the $\sigma$ field with itself. The coupling constants between 
$h^{(l)}$ and $\sigma$ are the constants $\gamma_{l}^{2}$ (\ref{6.35}) and the $h^{(l)}$
exchange graphs with external lines on the AdS boundary equal the corresponding 
CFT amplitudes for the exchange of the conserved current $J^{(l)}$ between  $\alpha$ 
fields
\begin{equation}
\frac{1}{2N} W_{\beta}(x_1,x_2,x_3,x_4;d-2+l, l) = \frac{1}{2N}\gamma_{l}^{2}
(x_{12}^{2}x_{34}^{2})^{-2} u^{-\frac {3}{2}}\sum_{n,m=0}^{\infty}
a_{nm}^{(l)}\frac{u^{n}(1-v)^{m}}{n!m!}
\label{7.3}
\end{equation}
Here $u, v$ where defined through $x_1,x_2,x_3,x_4$ by (\ref{5.5}) with labels (1,2,4,5) 
replaced by (1,3,2,4), and the channel is $(1,3)\leftrightarrow (2,4)$.  We remember the 
reader that the coefficients $a_{nm}^{(l)}$ are independent of the external lines of the 
exchange graph if these are all equal. Therefore for the external fields $\alpha$ we can 
use the same coefficients as for the external lines $\vec{\varphi}$  (\ref{6.34}).

The sum over $l$ can then be performed
\begin{equation}
\sum_{l~even >0}\gamma_{l}^{2}	= 4lim_{d\rightarrow3}(d-3)^{2}C_{nm}
\label{7.4}
\end{equation}
where $C_{nm}$ was given in \cite{5}, eqns. (23), (24). Obviously only the second order
poles in $d$ at $d=3$ are relevant. For this channel we obtain 
\begin{equation}
\frac{1}{2N}(x_{12}^{2}x_{34}^{2})^{-2}\{ (uv)^{-\frac{3}{2} }(1-u-v) - u^{-\frac{3}{2}}
(1+u-v)\}
\label{7.5}
\end{equation}
For the crossed channels we get instead
\begin{eqnarray}
(1,2)\leftrightarrow(3,4): \{\ldots\} &=&  -u^{-\frac{3}{2}}(1+u-v) -v^{-\frac{3}{2}}(1-u+v) \\
\label{7.6}
(1,4)\leftrightarrow(2,3): \{\ldots\} &=&  (uv)^{-\frac{3}{2}}(1-u-v) - v^{-\frac{3}{2}}
                                            (1-u+v)
\label{7.7}
\end{eqnarray}
Summing all channels we obtain for the $O(N^{-1})$ contribution to the four-point function
\begin{equation}
\frac{1}{N}(x_{12}^{2}x_{34}^{2})^{-2}\{(uv)^{-\frac{3}{2}}(1-u-v) - u^{-\frac{3}{2}}
(1+u-v) - v^{-\frac{3}{2}}(1-u+v)\}
\label{7.8}
\end{equation}

We know \cite {6} that this is the correct expression for the four-point function
of $\alpha$ in the minimal $O(N)$ sigma model. A quadruple local self-interaction of $\sigma$
does therefore not arise in HS(4) at this order. We know also \cite{6} that any n-point function of
$\alpha$ for even n in the $O(N)$ $\sigma$ model at $d=3$ can be obtained by insertion of the 
connected part of the four-point function (\ref{7.8}) and additional propagators at order 
$O(\frac{1}{N})$. If n is odd the n-point function is at least $O(N^{-\frac{3}{2}})$. This
carries over to HS(4) and the n-point functions of $\sigma$. In HS(4) only the A-terms 
of the exchange amplitudes must be considered. The tensorial form of the local coupling $\sigma\sigma
h^{(l)}$ is irrelevant, if only the traceless part of $h^{(l)}$ is taken into account \cite{17}.

It is obvious that to the perturbative order $O(N^{-1})$  a Lagrangian formulation of HS(4) with only 
local $\sigma\sigma h^{(l)}$ couplings of order $N^{-\frac{1}{2}}$ is possible. For further reading on HS(d+1)
models see \cite{18,19,20} and for the relation between HS(4) and the critical O(N) sigma model see \cite {21}.

\end{document}